\documentclass[twoside,11pt]{article}

\usepackage{jmlr2e}
\usepackage{booktabs} % For formal tables

\usepackage{subfig}
\usepackage[toc,page]{appendix}
\usepackage{color}
\usepackage{amsmath}
\usepackage[utf8]{inputenc} % allow utf-8 input
\usepackage[T1]{fontenc}    % use 8-bit T1 fonts
\usepackage{hyperref}       % hyperlinks
\usepackage{url}            % simple URL typesetting
\usepackage{booktabs}       % professional-quality tables
\usepackage{amsfonts}       % blackboard math symbols
\usepackage{nicefrac}       % compact symbols for 1/2, etc.
\usepackage{microtype}      % microtypography
\usepackage{amsmath}
\usepackage{amssymb}
\usepackage{graphicx}

\definecolor{mygreen1}{rgb}{0, 0.5, 0.1}

\newcommand{\se}[1]{{\textcolor{black}{#1}}}
\newcommand{\secor}[2]{{\textcolor{black}{#2}}}
\newcommand{\secmt}[1]{}

\newcommand{\ie}{\emph{i.e. }}

\definecolor{mypink1}{rgb}{0.858, 0.188, 0.478}

\definecolor{orange}{rgb}{1,0.5,0}

\begin{document}
\title{A multimodal movie review corpus for fine-grained opinion mining}
%\titlenote{Produces the permission block, and
  %copyright information}
%\subtitle{Extended Abstract}
%\subtitlenote{The full version of the author's guide is available as
%  \texttt{acmart.pdf} document}

\author{Alexandre Garcia, Slim Essid, Florence d'Alch\'e-Buc, Chlo\'e Clavel \\ T\'el\'ecom Paris, Universit\'e Paris-Saclay \\ \texttt{slim.essid,florence.dalche,chloe.clavel@telecom-paristech.fr}}

%\author{Alexandre Garcia}
%\affiliation{%
%  \institution{LTCI, Telecom ParisTech}
%  \streetaddress{46 rue Barrault}
%  \city{Paris}
%  \state{France}
%  \postcode{75013}
%}
%\email{algarcia@enst.fr}

%\author{Florence d'Alch\'e-Buc}
%\orcid{1234-5678-9011}
%\affiliation{%
%  \institution{LTCI, Telecom ParisTech}
%  \streetaddress{46 rue Barrault}
%  \city{Paris}
%  \state{France}
%  \postcode{75013}
%}
%\email{florence.dalche@telecom-paristech.fr}

%\author{Slim Essid}
%\orcid{1234-5678-9013}
%\affiliation{%
%  \institution{LTIC, Telecom ParisTech}
%  \streetaddress{46 rue Barrault}
%  \city{Paris}
%  \state{France}
%  \postcode{75013}
%}
%\email{slim.essid@telecom-paristech.fr}

%\author{Chlo\'e Clavel}
%\orcid{1234-5678-9014}
%\affiliation{%
%  \institution{LTCI, Telecom ParisTech}
%  \streetaddress{46 rue Barrault}
%  \city{Paris}
%  \state{France}
%  \postcode{75013}
%}
%\email{chloe.clavel@telecom-paristech.fr}

\maketitle

\begin{abstract}
In this paper, we introduce a set of opinion annotations for the POM movie review dataset, composed of 1000 videos. The annotation campaign is motivated by the development of a hierarchical opinion prediction framework allowing one to predict the different components of the opinions (\textit{e.g.} polarity and aspect) and to identify the corresponding textual spans. The resulting annotations have been gathered at two granularity levels: a coarse one (opinionated span) and a finer one (span of opinion components). We introduce specific categories in order to make the annotation of opinions easier for movie reviews. For example, some categories allow the discovery of user recommendation and preference in movie reviews. We provide a quantitative analysis of the annotations and report the inter-annotator agreement under the different levels of granularity. We provide thus the first set of ground-truth annotations which can be used for the task of fine-grained multimodal opinion prediction. We provide an analysis of the data gathered through an inter-annotator study and show that a linear structured predictor learns meaningful features even for the prediction of scarce labels. Both the annotations and the baseline system are made publicly available \url{https://github.com/eusip/POM/}.
\end{abstract}


\section{Introduction}

Due to the expansion of e-commerce on \secor{one side}{the one hand} and social networks on the other \secor{side}{hand}, opinionated contents have become available on a range of products going from purchasable goods (Amazon, PriceMinister reviews) to touristic services (Hotels and restaurants from TripAdvisor) and activities (Rotten Tomatoes, Imdb). Reviews often come under the form of a written commentary provided with one or more ratings summarizing the \secor{position of the reviewer toward}{reviewer's satisfaction-level with respect to} some aspects of the object being criticized. Even if these ratings provide a mean to measure the global satisfaction of customers, this information cannot be directly used to understand the specific aspects \secor{of their products they should put efforts on to improve the public response}{of the product which require improvement}. Different Machine Learning prediction tasks have been proposed to help understanding customer's satisfaction through the review they generate.

Prediction of the global polarity of a sentence or textual span, possibly with different intensity levels (varying from very negative to very positive), has been addressed as a sentiment analysis task in \cite{pang2002thumbs,maas2011learning,liu2012sentiment}. Following this path, different studies have built coding schema for describing and annotating further aspects of opinions. A common feature of these model is the definition of the functional components of an opinion and their properties (\textit{e.g.} implicit or explicit)  \cite{wiebe2005annotating}. The corresponding prediction task is commonly called \textit{fine-grained opinion mining} or \textit{Aspect-Based Sentiment Analysis} (ABSA) and consists in predicting the attitude of a speaker (named \textit{source}) toward an object (named \textit{target}).

The traditional approach developed in the ABSA task of the SEMEVAL campaign \cite{pontiki2016semeval} consists in two steps. First, the system has to identify whenever an opinion exists in a sentence, and to predict the corresponding target(s) \cite{jakob2010extracting}. Second, the polarity of each detected opinion expression is  computed \cite{wilson2005recognizing}. More recently, the task of fine-grained opinion mining has been cast as a structured prediction problem where the different components of an opinion are predicted at the same time \cite{garciaabstention,marcheggiani2014hierarchical}. These structured models take advantage of the relationship existing between the different components of an opinion to help predicting each of them. These methods rely on annotations gathered at the sentence level and possibly at the token level while at the same time using review level feedbacks such as star ratings. Unfortunately, the human interpretation of opinions expressed in the reviews is highly subjective and the opinion aspects and their related polarities are sometimes expressed in an ambiguous way and difficult to annotate \cite{clavel2016sentiment,marcheggiani2014hierarchical}. In the case of spoken language, this difficulty is even higher due to the lack of syntax of some sentences and the presence of disfluencies that break the continuity of the discourse.   %\al{Je vais essayer de trouver un exemple pour illustrer}. 

In this work, we propose flexible guidelines for the fine grained annotation of opinion structures in the context of video based movie reviews. The corresponding schema introduces some links between the coarse opinion recognition (at the review level) and the detection of token-level opinion functional components. This nested model ensures that the annotations are consistent at different levels of details and can be used in joint prediction models \cite{garciaabstention} to take into account the labeled information at each level. Since the working support of each annotator is a set of transcripts of spontaneous spoken reviews, the main difficulty is to provide guidelines that are flexible enough to match with the structure of oral language while ensuring a correct agreement between multiple workers. %\ch{je me permets de retirer la phrase suivante car 1) je ne suis pas d'accord 2) cela reduit complètement l' interet de ton annotation des le debut de l article }Whereas these annotations may be insufficiently accurate to be used as a ground truth for comparing different systems on the task of multimodal fine-grained opinion mining, we expect that they can be of interest as an intermediate representation for auxiliary tasks such as the \se{prediction of review-ratings} \cite{garciaabstention,marcheggiani2014hierarchical}. %and \al{citer plein de taches du challenge multimedia - dans partie suivante }. 

In Section 2, we present the previous studies concerning opinion annotation and especially the studies carried out on existing multimodal datasets. Then, we present the dataset we used (Section 3) and the protocol and the setting of our annotation campaign (Section 4). Finally, we present some results validating the dataset in Section 5. The annotations and the code required for running the experiments will be made publicly available. %In particular, previous work focusing on jointly predicting aspects and polarities mainly focused on text-based reviews in which extracting the relations between the components of an opinion is easier \cite{johansson2013relational,gangemi2014frame}. 

%\ch{Ici il faut presenter un peu plus TON approche hierarchique en referant a nouveau a ICML : To tackle this subjectivity issue, we develop a model able to learn structured outputs such as opinions and targets and to allow abstention on one of the components of the structure (\textit{e.g}. polarity or target)}
%\secmt{ca commence a etre long et on ne sait toujours pas ou on va. Avant d'entrer dans ces details, ce serait mieux juste apres le premier paragraphe de nous dire c'est quoi en gros l'objet du papier, ce que tu veux faire, la motivation, ce qui permet de comprendre le role de ce premier parag... et apres tu peux positionner}

%\ch{With the aim to extend our model to multimodal fine-grained opinion mining, we present here a coding scheme for the annotation of the opinion structure. The annotators were asked to identify opinions in an audiovisual dataset of movie reviews both at the textual span level and at the token level. They were given access to the manually detailed transcript of each video including the presence of vocal disfluencies such as the repetition of words during hesitations. }

\section{Related work}

The annotation of opinion in \se{natural} language is difficult due to the inherent subjectivity of the task and the need \secor{of}{for} a framework that ensures that different annotators work in a consistent way. An example of such a framework is the annotation scheme of the MPQA opinion corpus (news articles) \cite{wiebe2005annotating} which relies on the annotation of \textit{private state frames}, \ie textual spans that describe a mental state of the author. In the case of an opinion, it can describe either the target (what the \textit{private state} is about), the source or holder (who is expressing the opinion) and other characteristics such as polarity, intensity, attitude.
In \cite{toprak2010sentence}, the authors improve the annotation scheme  for consumer reviews by splitting it in two successive steps where the polarity and the relevance to the topic of the sentences is first examined and then the different opinion components are identified. They also go beyond the annotation of \textit{private state frames} and explicitly introduce some new labels: \textit{is reference} and \textit{modifiers} that link the different opinion components together. In this paper, we take a step in the same direction by proposing a fine-grained annotation of opinion components and we propose a new setting more flexible for the annotation of multimodal movie reviews.

Regarding multimodal review corpora, even though no fine-grained annotation of these datasets currently exist, %\ch{il existe peut etre des corpus de debats annotes en opinion, du coup j ai ajoute review}
different related annotation tasks have been proposed. Among these efforts, the ICT-MMMO corpus \cite{wollmer2013youtube} consists of 370 movie review videos for which an annotator has given an overall label: positive, negative or neutral, to describe the \secor{position}{viewpoint} of the reviewer. The recent CMU Multimodal SDK \cite{zadeh2018multi} provides a setting ready-to-use for building multimodal predictors based on opinionated or emotionally colored content. In the CMU-MISO dataset \cite{zadeh2016multimodal}, 93 videos have been gathered and annotated at the segment level in terms of intensity of the opinion expressed. In their case, opinion is defined as a \textit{subjective segment} for which a categorical label between 1 and 5 is given. This representation is in fact restrictive since it doesn't provide information about the target of the  expressed opinion. Besides, it doesn't provide information on the  cues that have been used in order to choose a particular intensity.   For the present first annotation campaign of fine-grained opinion in multimodal movie reviews, we use the Persuasive Opinion Multimedia (POM) dataset \cite{park2014computational} which consists of 1000 video-based movie reviews that were originally annotated in terms of persuasiveness of each speaker. In the next section, we present the different features of the POM database that led us to select it. % 

\section{The video opinion movie corpus}

\begin{figure}[h]
    \centering
    \subfloat[Negative opinion]{{\includegraphics[width=4cm]{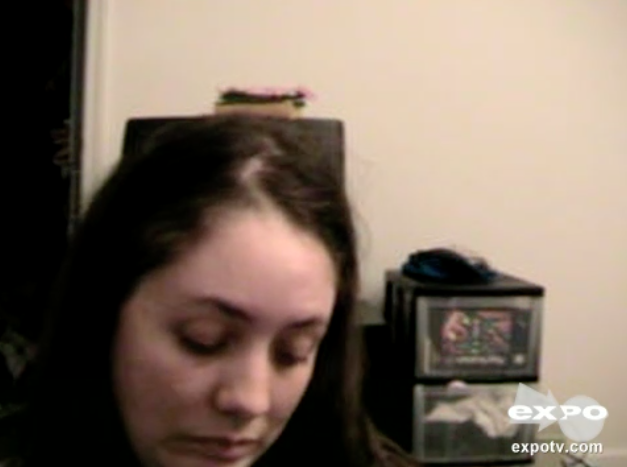} }}
    \subfloat[Other example of negative opinion]{{\includegraphics[width=4cm]{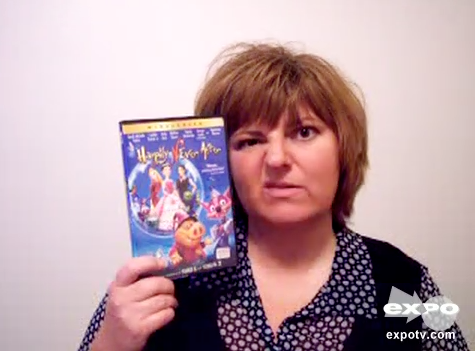} }}
    \qquad
    \subfloat[Positive opinion]{{\includegraphics[width=4cm]{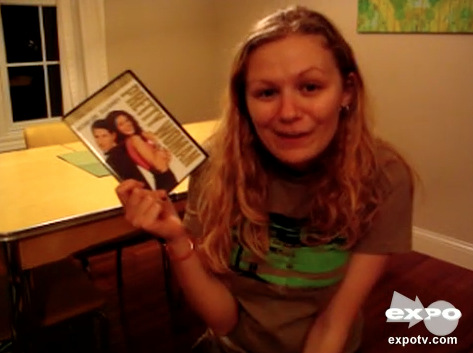} }}
    %\qquad
    \subfloat[Neutral opinion]{{\includegraphics[width=4cm]{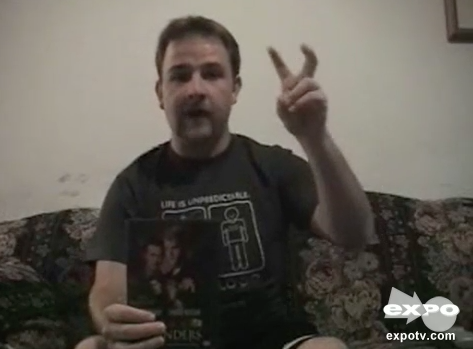} }}
    \caption{Examples of frames taken from different videos of the dataset illustrating the visual expression of opinions.}%
    \label{fig_1:example}%
\end{figure}

Our annotation campaign focuses on the identification of the opinions expressed in the POM dataset. In each video, a single speaker in frontal view gives his/her opinion on a movie that he/she has seen. The corpus contains 372 unique speakers and 600 unique movie titles. It has originally been built in order to analyze the persuasiveness of the speakers and no attention has been so far given to the content of the reviews themselves. We expect however that the use of multi-modal data can be of interest when predicting polarized content. Figure \ref{fig_1:example} shows examples where \secor{we can see that the frames}{it is clear that the visual content} may be crucial to disambiguate the polarity of some reviews (for example in the hard case of irony).

This dataset has been chosen for running an annotation campaign for the following reasons: 

\begin{itemize}
\item a defined setting for target identification: the documents contain opinionated content and are focused on a single type of target (here movie aspects) which makes it easier to build a typology of the possible targets for the target annotation task;
\item an illustration of spontaneous spoken expressions of opinions in a multimodal context: the reviews are based on spoken language for which the video is also available contrarily to previous studies of sentiment analysis based on phone call studies \cite{clavel2013spontaneous}. As a consequence, the annotation of the transcript is harder than for classical written language especially at a fine-grained level;% \ch{j ai rajoute classical car parfois dans les forums ou dans les tweets cela peut etre tres difficile a annoter }
\item a hierarchical representation of opinions: other auxiliary labels are available such as star ratings given by the reviewer, sentence-based summary and persuasiveness. The fine-grained annotations can be used as intermediate representations to help predicting these values \cite{garciaabstention}.
\end{itemize}

The POM dataset also provides a manual transcription for each review that we used in our annotation campaign. It contains 1000 reviews for which the average number of sentences per review is 15.1 and the average number of  tokens per sentence is 22.5. In its current version, this dataset only contains annotations performed at the review level. Indicators of the persuasiveness of the speaker are available (professionalism, quality of argumentation \ldots). Among the available data, the authors of \cite{park2014computational} asked the annotators to evaluate the polarity of the reviews by guessing its corresponding five-level star rating. The results in Table \ref{table:star_ratings} show that the reviews are strongly polarized which indicates the presence of clear opinion expressions. 

%\subsection{Statistics about the corpus }

% Since we base our annotation campaign on the transcripts of video based reviews, we first present an overview of the characteristics of such data. The 

% \begin{table}[!ht]
% \centering
% \caption{Statistics about the corpus \ch{Corpus statisctics}}
% \label{my-label}
% \begin{tabular}{|l|l|}
% \hline
% \textbf{}                & \textbf{Count in the dataset} \\ \hline
% Reviews                  & 1000                          \\ \hline
% Avg Sentences per review & 15.1                          \\ \hline
% Avg Tokens per sentence  & 22.5                          \\                      \\ \hline
% \end{tabular}
% \end{table}

\begin{table}[!ht]
\centering
\caption{Repartition of the star ratings at the review level}
\label{table:star_ratings}
\begin{tabular}{|l|l|l|l|l|l|}
\hline
Star rating                                                    & 1   & 2   & 3  & 4   & 5   \\ \hline
\begin{tabular}[c]{@{}l@{}}Number of\\ occurrences\end{tabular} & 253 & 200 & 61 & 133 & 353 \\ \hline
\end{tabular}
\end{table}

%\begin{figure}[!ht]
%    \centering
%\includegraphics[width=.4 \textwidth]{images/star_ratings.pdf}
%    \caption{Repartition of the star ratings at the review level}%
%    \label{fig_:star_ratings}%
%\end{figure}

% 1 : 252, 2:200, 3:61, 4:132, 5:353

In the next section, we detail our setup for the fine grained opinion annotation of the POM dataset.

% a termer mettre des stats sur les disfluences (pas priorite)
\section{Annotation}

\subsection{Opinion definition}

Following the path of previous opinion annotation studies \cite{langlet2017web} and based on appraisal theory \cite{martin2003language}, we define an opinion as the expression of a judgement of quality or value of an object. This definition makes it possible to represent an opinion (here called attitude) as an \textit{evaluation} (positive or negative) by a \textit{holder} (for example the person who expresses her opinion) of a \textit{target} (for example a service or a product). In the case of movie reviews, the opinion holder is the reviewer himself most of the time but some exceptions exist. For example, in the sentence \texttt{"my children like the characters of this cartoon"}, the \textit{holder} is 'children'. The \textit{target} component is defined here as a part of a hierarchically defined set of aspects \cite{Wei2010} which covers the subparts of the object examined (here movie reviews). Finally, the \textit{polarity} component indicates whether the evaluation is positive or negative. In what follows, we define an opinion as an expression for which these 3 components exist and are not ambiguous. The present definition does not include: i) emotions without any target %\al{citer} %\ch{je dirais plutôt emotion sans cible + donner un exemple.  je pense pense qu'un sentiment peut avoir une cible :}
\cite{Munezero2014AreTD} such as in the sentence "I was so scared",
%\ch{ distingue émotions, opinions et sentiments de cette manière : les émotions diffèrent des sentiments de par leur durée (les émotions sont des phénomènes plus brefs) et de par la présence d’une cible (les émotions ne sont pas toujours reliées à un objet). Les opinions sont des interprétations personnelles d’une information et ne sont pas nécessairement chargées émotionnellement comme les sentiments} 
and ii) polar facts \cite{jakob2010extracting} which denotes for facts that can be objectively verified but indirectly carry an evaluation such as in "What a surprise he plays the bad guy once again". In Section \ref{issues}, we provide guidelines to handle these cases in the annotation process.% and others presenting some annotation difficulties that we present in Section \ref{issues}

%old: Such an expression is defined by a \textit{holder} corresponding to the entity expressing his opinion. In the case of movie reviewed, this will be the viewer himself most of the time but some exceptions exist: in the case of cartoons, the reviewer can provide the opinion of his children which can be different from his own one. A second component is the \textit{target} corresponding to the object on which the opinion is expressed. This component can be decomposed on a hierarchically defined set of aspects \al{citer un truc} which covers different subparts of the object examined. Finally the \textit{polarity} component indicates whether the judgement is positive or negative. In what follows, we define as an opinion the expression for which these 3 components exist (even when they are implicit) and don't take into considerations sentiment expressions \al{citer} and polar facts \cite{jakob2010extracting}. However we provide guidelines to handle these cases and others presenting some annotation difficulties that we present in section \ref{issues}

\subsection{Fine-grained annotation strategy}
  We want to build a set of annotations that identifies the grounds on which the opinions of the reviewer are perceived by an annotator, both at the expression and at the token levels. We expect that better localizing the words which are responsible for the expression of an opinion may help finding the visual/audio features that carry the polarity information. Annotating this data is challenging due to the specific language structures of oral speech and the presence of disfluencies. We propose a two-level 
  %\ch{dire clairement quels sont ces deux niveaux sentence, expression, token? preciser des maintenant la distinction entre expression (que tu appelles span apres) et token, bien uniformiser les notations} 
  annotation method in order to (1) obtain a consistent identification of the opinion expressed in a sentence and the words responsible for this identification and (2) provide accessible guidelines to the annotators when the lack of grammatical structure of the sentences makes it difficult to find the delimitation of the phrases. For this second reason we define the expression level as 'the smallest span of words that contains all the words necessary for the recognition of an opinion'. These boundaries are in practice very flexible and might be very different from one annotator to the other. 

Once an opinion is identified at the expression level, the annotator is asked in a second time to highlight its different components based on the tokens located inside the previously chosen boundaries. In what follows, we refer to this step as the token-level annotation. It consists in selecting the group of tokens indicating the \textit{target}, \textit{polarity} and \textit{holder} of the opinion. In this case multiple spans can be responsible for the identification of each components. The instruction in such cases is to pick all the relevant spans for polarity tokens and only the most explicit one for target tokens. As an example in the sentence : \texttt{"It's the best movie I've seen"},  the selected polarity token  \secor{would be}{is} \textit{best}, the holder token is \textit{I} and the target token is \textit{movie} since it is more explicit than \textit{It}, \secor{that}{which} requires anaphora resolution to be understood.
%We provided some rules to the workers for such issues that we detail in the Section \ref{issues}

In the end, we provide a dataset with the following features :\\
$\bullet \ $ Span-level annotation :
\begin{itemize}
\item Opinion targets and polarities are annotated at the expression level. 
\item For each segment, the targets are categorized in a predefined set adapted to the context of movie reviews.
\item The corresponding polarities are then categorized on a five-level intensity scale.
\end{itemize}
$\bullet \ $ Token-level annotation :
\begin{itemize}
\item The words which led to the choice of the target category and polarity intensity are specifically annotated.
\end{itemize}
In the next section we study the difficulties specific to the corpus used.

\subsection{Annotation challenges and guidelines}
\label{issues}

We have previously highlighted the specificities of the dataset, namely the oral nature of the discourse and especially the presence of disfluencies and non grammatical phrases. For these reasons, defining precisely the textual span corresponding to an opinion is difficult. We tackled this issue by providing a rule of thumb to the annotators. Some difficulties remain, owing to the non professional nature of the movie reviews: not only do the reviewers give their opinion about the movie itself, but also they take into account the background of the viewer and tend to give some \secor{advices}{advice}. For this reason, the reviewers regularly give a \textit{recommendation} for the viewers that are likely to enjoy the movie being examined. In this case the opinion of the reviewer \secor{himself}{him/her-self} toward the movie is unclear, as it can be seen in the sentence: \texttt{"This movie is perfect for kids"}.
Consequently, we have asked the annotators to indicate whenever this type of sentence appears, in order to avoid adding the complexity of a dedicated treatment. This annotation takes the form of a boolean variable attached to an expression as it is shown in Figure \ref{fig_:schema}.

A second case is the comparison between the movie reviewed and the other ones such as the different elements of a saga or even related movies (such as movies with some actors in common or the same director). When this happens, a \textit{comparison} occurs and the choice of the target of the opinion becomes ambiguous in sentences such as :\texttt{"Obviously Harry Potter 1 is better than this one."}. Once again the \textit{comparison} label dedicated to \secor{handle}{handling} these cases is defined \secor{on}{in} \secor{the}{}\secmt{DELETE THE} Figure~\ref{fig_:schema}.

Finally, some sentences may contain some polarized content \secor{and help understanding the attitude of the reviewer but do not admit}{conveying the attitude of the reviewer without holding} an explicit target. Other may have no target at all when they consist \secor{in}{of} a sentiment expression. Such sentences have been referred to in previous work as \textit{Speaker’s emotional state} \cite{mohammad2016practical} or \textit{polar fact} \cite{jakob2010extracting}. Since these sentences are hard to annotate (both in terms of target choice and boundary selection) we ask the annotators to specifically identify them using the \textit{sentiment} tag. This enables us to separately treat the sentences in which the target is known but does not appear, as for example in \texttt{"I must say that what I heard sounded good."} where the target is obviously the music even if its not stated, and the sentences in which the target is really ambiguous or inexistent. 

These \secor{3}{three} labels are incorporated in the annotation tool under the form of boolean variables tied to the span level annotation that can be selected. When at least one of the 3 labels $\{$ \textit{recommandation}, \textit{comparison}, \textit{sentiment} $\}$ is active, we do not ask the annotators to perform the second step of token-level annotation since we do not consider these spans as real opinions. 

%inspirer de l article de Cardie - partir d exemples et notamment les exemples du guide

%specifique a l oral
%disfluencies 

%specifique a la critique 
%\al{Ajout de Recommandation, comparison, no target (sentiment) } No-opinion s 

\subsection{Annotation schema}

The annotation campaign has been run on a remotely hosted platform running the Webanno tool \cite{de2016web}. This choice was motivated by the simplicity of the configuration of multiple tag layers and \secor{online configuration}{the possibility of performing this configuration online}. When logged into the platform, an annotator can select a transcript of a movie review assigned to him/her and each annotation added is automatically saved.

%\ch{The Webanno tool ... decrire les atouts de l'outil : tourne sur une plateforme internet et permet de delimiter des frontieres de span et de poser un label dessus de maniere ergonomique} 

\begin{figure}[!ht]
    \centering
\includegraphics[width=.4 \textwidth]{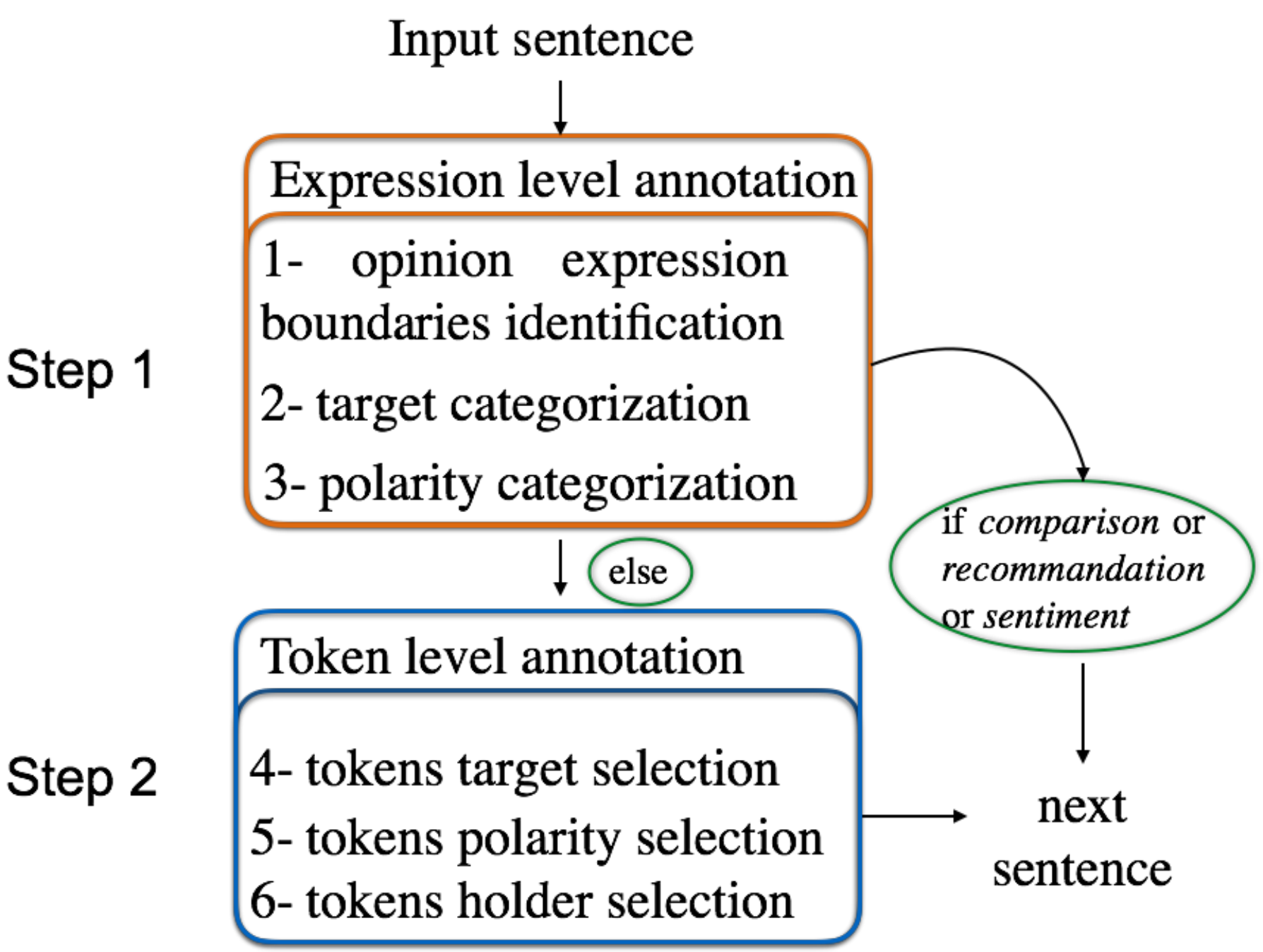}
    \caption{Annotation schema}%
    \label{fig_:schema}%
\end{figure}

The annotation task is split in two consecutive \se{subtasks} described in Figure~\ref{fig_:schema}.

%For each transcript the worker is asked to successively identify: 

%\begin{enumerate}
%\item The spans containing an opinion and the corresponding target .
%\item The polarity on a 1 to 5 scale of the corresponding opinions.
%\item The specific words defining the target.
%\item The specific words indicating the polarity.
%\item The words indicating the holder\secmt{define} when he/she is different from the speaker.
%\end{enumerate}

%\ch{presenter l arbre d annotation}

We additionally asked the annotators to identify the name of the movie reviewed when available.

The scheme is a coarse-to-fine annotation where the worker has to successively identify the textual spans containing an opinion; identify the corresponding target, then the polarity; and finally select the words that guided his/her choice. The possible labels for the categorization tasks are defined in advance: 

The taxonomy of targets is derived from the one of \cite{zhuang2006movie} and corresponds to the hierarchy reported in Table~\ref{set-targets}. Once the target is identified, the corresponding polarity is also chosen on a \secor{5 levels}{five-level} scale, from very negative to very positive.

\begin{table}[!ht]
\centering
\caption{Predefined targets for movie review opinion annotation}
\label{set-targets}
\begin{tabular}{|l|l|l|}
\hline
\textbf{Movie Elements}    & \textbf{Movie People}                 & \textbf{Support} \\ \hline
Overall                    & Producer                              & Price            \\ \hline
Screenplay                 & Actor - Actress                       & Availability     \\ \hline
Character design           &  \begin{tabular}[c]{@{}l@{}}Composer - singer\\ soundmaker\end{tabular}       & Other            \\ \hline
\begin{tabular}[c]{@{}l@{}}Vision and\\ special effects\end{tabular} & Director                              &                  \\ \hline
\begin{tabular}[c]{@{}l@{}}Music and\\ sound effects\end{tabular}    &  \begin{tabular}[c]{@{}l@{}}Other people involved\\ in movie making\end{tabular} &                  \\ \hline
\begin{tabular}[c]{@{}l@{}}Atmosphere\\  and mood making\end{tabular}
        &                                       &                  \\ \hline
\end{tabular}
\end{table}

In the context of this paper, we will only report results concerning \textit{Movie Elements} to focus the discussion on a reduced set of labels. The results concerning \textit{Movie People} and \textit{Support} will be provided on the page of the dataset. 

We detail the experimental protocol in the next section.

\subsection{Protocol}

We provided examples of annotated reviews in the \se{annotation} guide and trained three recruited workers on 150 reviews before beginning the annotation campaign. Then each of the 850 remaining reviews \secor{has been}{was} annotated once by one of the workers. Each annotator was given an access on a remotely hosted Webanno server where he/she could log him/her-self and annotate the transcripts of the review via a parameterized interface. Note that due to the explicitness of the reviews, we only provided the transcripts of the videos to each annotator which did not have to watch the videos (but were aware of the oral nature of the original content). 
An example of annotated review provided as an example in the annotation guide is given below in Figure \ref{fig_2:example_annotation}:

\begin{figure}[!h]
    \centering
\includegraphics[width=.5 \textwidth]{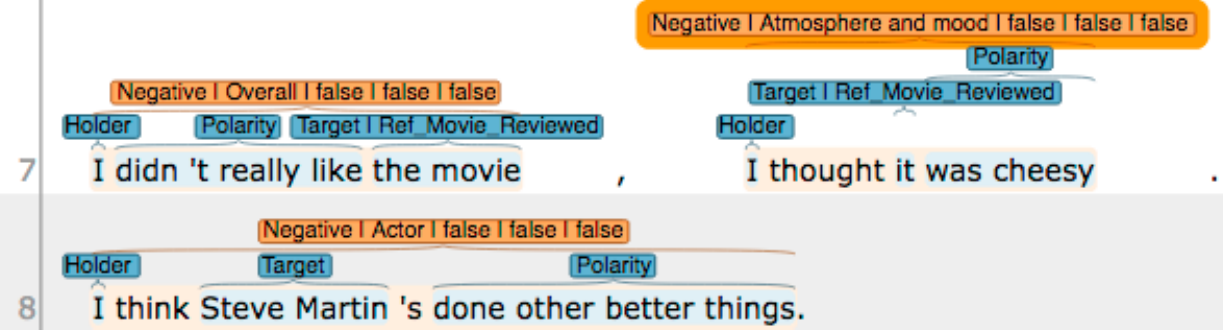}
    \caption{Extract from the annotation of the review of the movie : Cheaper by the Dozen }%
    \label{fig_2:example_annotation}%
\end{figure}

Since the tasks have been shared among different workers, an issue is the variability of the annotations. In the next section we focus on issues raised by the multi-annotator setting.

%\section{Statistics about the corpus}
%M\ch{Annotation overview}
%\subsection{Annotation distribtion in the whole corpus}

%Pt1 : stats sur le texte

%\begin{table}[!h]
%\centering
%\caption{Statistics about the corpus \ch{Corpus statisctics }}
%\label{my-label}
%\begin{tabular}{|l|l|}
%\hline

%Avg opinions per review  & 6,1   \ch{a mettre en 3.5 dans le protocole}                        \\ \hline
%\end{tabular}
%\end{table}

%\al{stat utilisable : Avg opinions per review = 6.1}

%Pt 2 : stats sur les annotations : 

%\begin{figure}[!h]
%    \centering
%    \subfloat[polarity of the opinions]{{\includegraphics[width=4cm]{images/polarities_hist_span.pdf} }}
%    \subfloat[movie element targets count]{{\includegraphics[width=4cm]{images/targets_elem_hist_span.pdf} }}
%    \qquad
%    \subfloat[Polarity spans identified at the token level]{{\includegraphics[width=4cm]{images/polarities_hist_token.pdf} }}
    %\qquad
%    \subfloat[movie people targets count]{{\includegraphics[width=4cm]{images/targets_char_hist_span.pdf} }}
%    \caption{Bar plot of the different annotations \ch{voir comment mieux presenter (on ne voit pas)}}%
%    \label{fig_n:histograms}%
%\end{figure}

%\ch{commenter la figure + ajouter la longueur moyenne d un span d opinions, d un span de target, d un span de polraite (positif negatif), d un span de source}

\section{Validation of the annotation}
\label{validation}

We examine the quality of the annotation by two means: \se{using} a measure of the inter-annotator agreement on a data subset; and \se{performing} a study of the most influential linguistic features used by a structured linear model on the whole annotated corpus.

\subsection{Inter-annotator agreement}
We measure the inter-annotator agreement by computing the \secor{kappa cohen}{Cohen's kappa coefficient} on two groups of 25 reviews that were annotated by two different annotators. We only gathered double annotations on a small subset of the dataset for annotation cost reasons. Since we are in a multilabel setting (spans of different opinions can overlap), we compute an agreement for each label: The compared objects are binary sequences labeled as 1 if the label is active and 0 otherwise. We denote by the letter A (resp. B), the reviews annotated by reviewer 1 and 3 (resp. 2 and 3) and report the results at the span and sentence level in Table \ref{kappas} and Table \ref{kappas_pol}.

%\ch{rappeler  qu' au niveau du span et au niveau de la phrase tu es dans un cas multilabel, et expliquer que comme tu es dans le cas multilabel  tu calcules un kappa indépendamment pour chaque label en consderant a chaque fois deux categories exclusives : label-nolabel}

%During the annotation campaign, 

%\ch{kappa et commenter }
%\al{A: batch 11, B : batch 20}

%\ch{ajouter ici des elements sur les accords interannotateurs : calculer un kappa global }

%1)  Target kappas
\begin{table}[!ht]
\centering
\caption{Cohen's kappa at the span and sentence level for the target annotations and total number of segments annotated by the two workers}
\label{kappas}
\begin{tabular}{lrrrr}
\toprule
{} &  A\_span &  B\_span &  A\_sent &  B\_sent \\
\midrule
\begin{tabular}[c]{@{}l@{}}Atmosphere\\ and mood\end{tabular}
        &       0.00 (69) &      0.00 (149) &       0.00 (12) &      -0.01 (19) \\
\begin{tabular}[c]{@{}l@{}} Character\\ design \end{tabular}
        &       0.00 (12) &      0.00 (33) &       0.00 (1) &      0.00 (4) \\
\begin{tabular}[c]{@{}l@{}}Music and\\ Sound effects\end{tabular}
    &       0.48 (78) &            - (0)&       0.57 (7) &            - (0)\\
Overall                   &       0.32 (1818)&       0.46 (2268)&       0.41 (188)&       0.55 (201)\\
Screenplay                &       0.23 (194) &       0.14 (187)&       0.23 (16)&       0.32 (18)\\
\begin{tabular}[c]{@{}l@{}}Vision and\\ Special effect\end{tabular}
  &       0.08 (120) &       0.32 (43) &       0.25 (8)&       0.50 (4)\\
\bottomrule
\end{tabular}
\end{table}

%2) polarity kappas

\begin{table}[!ht]
\centering
\caption{Cohen's kappa at the span and sentence level for the polarity annotations and total number of segments annotated by the two workers}
\label{kappas_pol}

\begin{tabular}{lrrrr}
\toprule
{} &  A\_span &  B\_span &  A\_sent &  B\_sent \\
\midrule
Negative         &           0.30 (928) &           0.41 (1210)&           0.51 (106) &           0.64 (141)\\
Positive         &           0.22 (675)&           0.34 (792)&           0.59 (145)&           0.55 (83)\\
\begin{tabular}[c]{@{}l@{}} Mixed -\\ Neutral \end{tabular}    &           0.00 (47)&           0.40 (205)&          0.00 (5)&           0.53 (22) \\
\midrule
\begin{tabular}[c]{@{}l@{}} Opinion\\ presence \end{tabular} &           0.37 (1650) &           0.52 (2207)&           0.44 (256)&           0.58 (246)\\
\bottomrule
\end{tabular}

\end{table}

We additionally defined a global kappa which indicates the confidence with which an opinion can be recognized. We refer to this table as \textit{Opinion presence}. The corresponding obtained kappas refer to  moderate agreement \cite{landis1977measurement}, which is very encouraging for subjective phenomena such as opinions. 

Regarding the target, the repartition of labels is imbalanced : the \textit{Overall} label is strongly dominant whereas \textit{Character design} or \textit{Music and Sound effects} are very rare. Drawing some conclusions on the rare labels is impossible but we still observe that some moderate agreement can be measured for the overall class at the sentence level.

Concerning the polarity annotations, the labels are slightly better balanced leading to higher confidence in the results. %divided between the labels. 
Relaxing the annotation at the sentence level raises the agreement from low to moderate which indicates that the low results at the span level are implied by the absence of hard annotation guidelines for the identification of the span frontiers.

%\textit{Atmosphere and mood} and \textit{Character design} are systematically close to 0. Since these labels are very rare, the size of the sample is not sufficient to conclude on the values. We observe a similar behavior for \textit{Music and sound effects} and \textit{Vision and Special effect} where the value are very different on the group A and B, indicating a strong variability due to the size of the sample. 

%\ch{ajouter kappa pour presence d une opinion vs absence (et on peut espérer obtenir dans ce cas un bon kappa}

%\ch{Il faut ajouter ici une discussion qui explique pourquoi les kappas sons particulierement bas. Tu peux te baser sur l echelle de Landis and Koch \url{https://fr.wikipedia.org/wiki/Kappa_de_Cohen}. On obtient quand meme des niveaux d accord moderes pour certains labels ce qui est rassurant }

\subsection{Study of linguistic features using a CRF-based model}

Since the results provided by the previous measures of inter-annotator agreement are not relevant for rare labels due to the size of the used sample, we additionally train a linear structured prediction  model for the task of opinion classification both at the token and the sentence level. By taking as input features the tokens themselves, we show that the learned model focuses on relevant vocabulary even for rare labels.% and \ch{derniere partie de phrase a retirer }contradicts the low agreement measure \al{Mouais}.

%\al{Dans la veine de : \cite{jakob2010extracting}}

%\ch{Suite a mettre a jour} In this section we propose a baseline for fine-grained opinion mining on the multimodal movie review corpus. 
We first consider the task of aspect and polarity prediction based on the span level annotations: we take as input features the sum of the one hot encoding of each word and the ones situated in a 5-token window. Each output object is a sequence of labels (one per token) corresponding to the span-level annotation previously described. Next, we treat the same task at the sentence level. The input features consist of the sum of the one-hot encoding of each token in the sentence and the output representation is built in the following way : we omit the polarity intensity information and introduce a \textit{Mixed} class indicating whether a sentence contains both positive and negative opinions. We also include sentences containing only neutral opinions in this class. Otherwise if the sentence contains at least one positive (respectively negative) opinion it is labeled as positive (respectively negative). 

A linear Conditional Random Field (CRF) \cite{lafferty2001conditional} model is trained for each label using the python-crfsuite library\footnote{\url{https://python-crfsuite.readthedocs.io/en/latest/}}.
We discarded the 150 texts used for the training of the annotators and split the remaining 850 texts in 5 folds. We tuned the parameters to optimize the macro-f1 score by cross-validation. We report the F1 score for each label averaged over the 5 folds in Table \ref{f1_polarity}.   %\al{Rajouter des trucs rapides à faire} 

\begin{table}[!ht]
\centering
\caption{F1 score for token and sentence level polarity prediction and corresponding number of occurrences in the dataset}
\label{f1_polarity}
\begin{tabular}{|l|l|l|}
\hline
            & Sentence level & Span level \\ \hline
Positive    & 0.67   (2218)        & 0.39 (26071)      \\ \hline
Negative    & 0.56    (1795)       & 0.26 (22988)      \\ \hline
Mixed       & 0.11  (299)         &           \\ \hline
No polarity & 0.87  (8737)         & 0.92  (243850)     \\ \hline
\end{tabular}
\end{table}

The reported scores are obtained both at the token level and at the sentence level. A crucial aspect is the dependency in the number of examples of each label. The results obtained for rare labels such as \textit{Mixed} is high precision / low recall. This behavior is due to the presence of specific vocabularies for which the predictor is guaranteed to accurately predict the polarity. We can display the vocabulary on which the model makes its prediction by analyzing the weights learned by our model. Let $\mathbf{x}=\{x_1,\ldots,x_T\},\mathbf{y} = \{y_1,\ldots,y_T\}$ two sequences of vectors of length $T$, $\forall t \in \{1,\ldots,T\} \; x_t \in \mathbb{R}^p \; y_t \in \mathbb{R}^q$. A linear chain conditional random field parameterizes the conditional distribution $p(\mathbf{y}|\mathbf{x})$ under the form :
\begin{equation}
p(\mathbf{y}|\mathbf{x}) = \frac{1}{Z(\mathbf{x})} \prod_{t=1}^T \exp\{\sum_{k=1}^K \theta_k f_k(y_t,y_{t-1},\mathbf{x}_t)\}
\end{equation}
Where $\mathbf{\theta}$ is the vector of learned weights, $Z(\mathbf{x})$ is an input dependent normalization term and $f_k, k\in \{1,\ldots,K\}$ is a set of feature functions. In the setting described here, these feature functions can be grouped in two categories : (i) output-output feature functions $f_k(y_t,y_{t-1},\mathbf{x}_t) = f_{k-(oo)}(y_t,y_{t-1})$ that do not depend on input data and (ii) input-output feature functions $f_k(y_t,y_{t-1},\mathbf{x}_t) = f_{k-(io)}(y_t,x_t)$ of the form :
\begin{equation}
f_{k-(io)}(y_t,x_t) = \begin{cases} 1 \; \text{if} \; y_t = y_k, \; x_t = x_k\\ 0 \;\text{else}
\end{cases}
\end{equation}
We only consider input-output feature functions and report the couples $(x_k,y_k)$ with highest weights $\theta_k$ in the Table \ref{pol_features}. We consider these weights as \textit{scores} since couples with higher $\theta_k$ values tend to increase the likelihood of the sequence  $p(\mathbf{y}|\mathbf{x})$.

%The best scored vocabulary\secmtt{Tu dois expliquer ce que ca veut dire : features with highest weights... ce n'est pas si evident} for each model is displayed in Table \ref{pol_features}:

\begin{table}[!ht]
\centering
\caption{Highest score input features for polarity label prediction at the sentence and the token level}
\label{pol_features}
\begin{tabular}{|l|l|l|}
\hline
\begin{tabular}[c]{@{}l@{}}Corresponding \\ label\end{tabular} & Sentence level                                                                        & Span level                                                       \\ \hline
Mixed                                                          & 'okay', 'average'                                                                     &                                                                   \\ \hline
Positive                                                       & \begin{tabular}[c]{@{}l@{}}'hilarious','amazing',\\ 'fantastic'\end{tabular}          & \begin{tabular}[c]{@{}l@{}}'great','cool',\\ 'good'\end{tabular}  \\ \hline
Negative                                                       & \begin{tabular}[c]{@{}l@{}}'disappointing',\\ 'disappointed',\\ 'boring'\end{tabular} & \begin{tabular}[c]{@{}l@{}}'terrible','not',\\ 'bad'\end{tabular} \\ \hline
No polarity                                                              & 'Thanks','Thank','review'                                                             & \begin{tabular}[c]{@{}l@{}}punctuation,\\ but','and'\end{tabular} \\ \hline
\end{tabular}
\end{table}

The top scored vocabulary raises two remarks :  
\begin{itemize} 
\item The polarized sentences - spans are mainly recognized through evaluative adjectives which are obviously linked to the corresponding label. 
\item The absence of polarity is treated in a different way at the sentence and span level. 
\end{itemize}
At the sentence level, the absence of polarity is systematic in sentences that introduce or conclude the review. The displayed vocabulary is characteristic of concluding sentences. At the span level, the punctuation and conjunctions which separate different opinions play an important role. These tokens receive a high score since they appear specifically at the boundary of an opinion. %ca compense le score des transition features. 

Finally we train a model for sentence-level target prediction and report the results in \secor{the}{}Table~\ref{target_sentence_voc}:

\begin{table}[!ht]
\centering
\caption{Highest score input features for aspect prediction at the sentence level}
\label{target_sentence_voc}
\begin{tabular}{|l|l|}
\hline
\begin{tabular}[c]{@{}l@{}}Corresponding \\ label (F1 score)\end{tabular}       & Highest score tokens                                                                                                                                 \\ \hline
Overall  (0.63)                                                            & \begin{tabular}[c]{@{}l@{}}'boring','disappointing','great','awesome',\\ 'terrible','wonderful','horrible'%,'disappointment'
\end{tabular}      \\ \hline
%Screenplay (0.35) 
\begin{tabular}[c]{@{}l@{}}Screenplay\\ (0.35)\end{tabular}
& \begin{tabular}[c]{@{}l@{}}'plot','storyline','story','interesting','nowhere',\\ 'predictable','stupid','slow','cheesy','dialogue'\end{tabular}      \\ \hline
\begin{tabular}[c]{@{}l@{}l@{}}Vision and \\ Special \\ effect (0.26) \end{tabular} & \begin{tabular}[c]{@{}l@{}}'beautifully','animation','effects',\\ 'cinematography','visually','graphics','picture'\end{tabular} \\ \hline

\begin{tabular}[c]{@{}l@{}l@{}}Music and \\ Sound \\ effects (0.04) \end{tabular}
                                               & \begin{tabular}[c]{@{}l@{}}'soundtrack','music','bands','quality','song',\\ 'sound','good','great','Oskar','songs'\end{tabular}                      \\ \hline

\begin{tabular}[c]{@{}l@{}}Character  \\ design (0.09)  \end{tabular}
                                                     & \begin{tabular}[c]{@{}l@{}}'characters','character','oh','awful','portrayal',\\ 'He','spinning','wonderful','good','running'\end{tabular}            \\ \hline

\begin{tabular}[c]{@{}l@{}}Atmosphere  \\ and mood (0.35) \end{tabular}
                                                  & \begin{tabular}[c]{@{}l@{}}'funny', 'fun', 'hilarious','funniest', 'cheesy',\\ 'Will', 'laughing', 'saying', 'many'\end{tabular}                           \\ \hline
\end{tabular}
\end{table}

Once again the low results are characterized by a low recall: a few words characterizing the presence of the target appear in the top score vocabulary but as the score decreases, some non characteristic words are quickly raised ('good', 'great' for \textit{Music and Sound effects}, 'oh', 'awful', 'He' for \textit{Character design}). These labels are specifically hard to predict due to the diversity of the vocabulary implied and the low number of examples available. 

The \textit{Overall} category is characterized by polarity words only. This is coherent with our annotation instructions : An opinion is labeled as \textit{Overall} if it targets the overall movie or if no category in the proposed hierarchy fits the opinion expressed. As a consequence the \textit{Overall} opinion is characterized by polarity words indicating an opinion but do not indicate a specific aspect.

\section{Conclusion and perspectives}

In this paper, we have presented the protocol and results of a fine-grained opinion annotation campaign for spoken language, based on a multimodal movie review dataset. The resulting annotations show low inter-annotator agreements at the token level but achieve better values by relaxing the annotation \secor{grain}{granularity}, placing it at the sentence level. Besides, the linear structured predictor learns meaningful features even for the prediction of scarce labels.  In future work, we plan to jointly use the different levels of granularity (from the review to the token level) in a hierarchical prediction framework in order to increase the accuracy of the predictor on each individual task.

%\begin{acks}
%\secmtt{Enlever ca pour l'instant : ca peut etre considere comme violant la contrainte d'anonymite, motif de rejet sans review !! ca nous est deja arrive}
%We are grateful to the MultiComp Lab from Carnegie Mellon University for sharing the POM dataset since its creation date. %The annotation campaign was funded by the T\'el\'ecom ParisTech Chair on Machine Learning for Big Data.

%\end{acks}

%\appendix
%Appendix A
%\section{If necessary}
%\ch{remercier l equipe de Morency qui nous a mis a disposition le corpus des sa creation}

\bibliographystyle{ACM-Reference-Format}
\bibliography{sample-bibliography}

\end{document}